\documentclass[conference,10pt]{IEEEtran}
\normalsize
\usepackage{graphicx}
\usepackage{amssymb,amsmath}
\usepackage{epstopdf}
\usepackage{amsthm}
\usepackage{color}
\usepackage{amsfonts}
\usepackage{amsfonts}
\usepackage{cite}
\usepackage{algorithmic}
\usepackage[center]{caption}
\usepackage{epsfig}
\usepackage{latexsym}
\usepackage{epstopdf}
\usepackage{verbatim}
\usepackage{amsthm}
\usepackage[linesnumbered,ruled,vlined]{algorithm2e}

\graphicspath{{./figures/}}

\begin{document}
%
\title{On the Degrees of Freedom of SISO X-Channel with Alternating CSIT}

\author{\large Ahmed Wagdy$^\star$, Amr El-Keyi$^\dagger$, Tamer Khattab$^\star$, Mohammed Nafie$^\dagger$ \\ [.1in]
\small
\begin{tabular}{c} $^\dagger$Wireless Intelligent
Networks Center (WINC), Nile University, Smart Village, Egypt.\\

$^\star$Electrical Engineering, Qatar University, Doha, Qatar.\\
\end{tabular}
}

\maketitle
\begin{abstract}
In this paper, we establish the degrees of freedom (DoF) of
the two-user single input single output (SISO) X-channel   
with alternating channel state information at the transmitters
(CSIT).
Three cases are considered for the availability of CSIT; perfect,
delayed and no-CSIT. Each state is associated with a fraction of time 
denoted by $\lambda_P, \lambda_D$ and $\lambda_N $, respectively.
We provide new results for DoF of the two-user SISO X-channel when the available CSIT
alternates between these three cases under a certain distribution $\Lambda(\lambda_P, \lambda_D, \lambda_N )$. Specifically, we
show that the X-channel with alternating CSIT for $\Lambda(1/8, 3/8, 1/2)$ can
achieve $5/4$ DoF. The interesting thing about $5/4$ is that it represents
a position of compromise or a middle ground between the channel knowledge that transmitters need to steer interference and the degrees of freedom that the network can achieve. Moreover, $5/4$ is strictly greater than $6/5$ which is the upper bound for the X-channel with \textit{fixed} delayed CSIT. 
\end{abstract}

\IEEEpeerreviewmaketitle

\section{Introduction}

The number one enemy and the bottle neck of wireless networks is
signal interference caused by the inherent broadcast nature of the
wireless medium.
Interference alignment (IA) \cite{maddah2008communication,4567443} is a key concept that arises in
the context of interference management which refers to creating a
correlation or an overlap between the interference signals at
receiver in order to minimize dimensions of the interference
subspace and maximize the desired signal space dimensions. Global,
perfect, and instantaneous channel state information at
transmitter (CSIT) is a fundamental component for interference
alignment in pioneering works
\cite{maddah2008communication,4567443,5208535,
 4418479}. On the other hand, in the complete absence of CSIT, an overly pessimistic assumption, the potential of IA fades and the degrees of freedom (DoF) of many networks collapse to what is achievable by time-division
between users\cite{6205390}.


Maddah Ali and Tse \cite{maddah2012completely} proved that completely outdated CSIT
can still be useful even if the channel states are completely
independent, contrary to the popular belief that in fast fading
environment, delayed CSIT is vain. 
In another research direction, Tandon et al. in \cite{6471826}
formalized an interesting model for the availability of the CSIT, 
in the context of broadcast channel,
called alternating CSIT.
In this model, the authors allow the availability of CSIT to vary
over time which is a more practical assumption and convenient to
the nature of wireless networks and alternating CSIT could
provide synergistic DoF gain. 

Earlier research works on the DoF of
the X-channel have determined that the upper bound for
DoF of a single-input single-output (SISO) X-channel is
$4/3$ and for a MIMO one is $4M/3$ where $M$ is the number of
antennas per node \cite{maddah2008communication, 5208535,4418479}.
These upper bounds are achievable with global, perfect and
instantaneous CSIT.
Then, the authors of \cite{ghasemi2011degrees} demonstrated that the delayed CSIT 
is beneficial for wireless networks consisting of distributed 
transmitters and receivers. Specifically, they showed that the SISO
X-channel with delayed CSIT can achieve $6/5$ DoF. Recently, the authors of
\cite{Delayed_CSIT} confirmed that the upper bound on the DoF of the SISO X-channel   
with delayed CSIT is to be $6/5$, using a new converse proof. 
More recently, the remarkable finding of \cite{Wagdy} is that the \textit{fixed} 
perfect CSIT is not a necessary requirement to achieve the upper 
bound on the DoF of the X-channel. 

In this work, we consider the SISO X-channel with
alternating CSIT. The main contribution of this work is
to provide new achievability results on the DoF of X-channel. 
A pertinent question we ask is whether the \textit{fixed} 
delayed availability of CSIT is the only middle ground between the 
two CSIT extremes; fixed perfect and fixed no-CIST.
We answer this question by
extending the system model of \cite{Wagdy} where each CSIT availability state
is associated with a time fraction $\lambda_S$ where $S \in \{P , D, N \}$ and each
CSIT alternation pattern is associated with a distribution
$\Lambda(\lambda_P, \lambda_D, \lambda_N )$ to the states of CSIT. 
Hence, the fixed delayed CSIT $\equiv \Lambda(0, 1, 0)$ is not the only middle ground between the fixed perfect $\equiv\Lambda(1, 0, 0)$ 
and the fixed no-CSIT $\equiv\Lambda(0, 0, 1)$. 
Finally, we show that for a certain distribution of CSIT availability states there are synergistic alternation patterns and
dissociative ones.

The remainder of this paper is organized as follows. In Section II,
the system model is described. Section III presents achievability scheme for three
different cases of CSIT alternation patterns.
Finally, Theorem 1 and concluding remarks are presented in Section IV.

\section{System Model}
A SISO X-channel is considered, in which a two
independent transmitters $T_1$ and $T_2$ transmit four
independent messages $W_{11}, W_{12},W_{21}, W_{22}$ to receivers
$R_1$ and $R_2$, where $W_{ij}$ originates in transmitter $j$ and
is intended to receiver $i$ and each node equipped with single
antenna. The received signal at the $i$th receiver at time slot
$t$ is given by:
\begin{equation}\label{eq1}
Y_i(t)= \displaystyle\sum_{j=1}^2 h_{ij}(t)X_j(t)+N_i(t)
\end{equation}
where $X_j(t)=f_{1j}(t) W_{1j}(t)+f_{2j}(t) W_{2j}(t)$ is the
transmitted signal from $T_j$ at the $t$th time slot which
satisfies the power constraint $E[\|X(t)\|^2]\leq P$ and
$f_{ij}(t)$ is the precoding coefficient for the message
$W_{ij}(t)$. The noise $N_i(t)\sim \mathbb{C}\mathcal{N}(0,1)$ is
the circularly symmetric complex additive white Gaussian noise
with unit variance generated at $R_i$ at time slot $t$. In
\eqref{eq1}, $h_{ij}(t)$ is the channel coefficient from $T_j$ to
$R_i$ and all channel coefficients are independent identically
distributed (i.i.d.) drawn from continuous distribution over time.

We assume that the receivers have perfect global channel state
information. Furthermore, we consider three different states of
the availability of CSIT identified by:
\begin{enumerate}
  \item Perfect CSIT (P): identifies the state of CSIT in which CSIT is available to the transmitters instantaneously and without error.
  \item Delayed CSIT (D): identifies the state of CSIT in which CSIT is available to the transmitters with some delay $\ge$ one time slot and without error.
  \item No CSIT (N): identifies the state of CSIT in which CSIT is not available to the transmitters at all.
\end{enumerate}

Then, The state of CSIT availability of the channels to the $i$th
receiver is denoted by $S_i$; where, $S_i$ $\in \{P,D,N\}$. In
addition, let $S_{12}$ denote the state of CSIT availability for
the channels to the first and second receivers, respectively.
Therefore, $S_{12} \in \{PP,PD,PN,DP,DD,DN,NP,ND,NN\}$. For
example, $S_{12}= PN$ refers to the case where $T_1$ has perfect
knowledge of $h_{11}$ (and no information about $h_{21}$) and
$T_2$ has perfect knowledge of $h_{12}$ (and no information about
$h_{22}$).

We denote the CSIT availability of the channels to the $i$th
receiver in the $t$th time slot by $S_i(t)$ and over $n$ time slots of time channel extension by $n$-tuple $S_i^n =(S_i(1),...,S_i(n))$.
Similarly, we denote the availability of CSIT for the channels to
the first and second receivers in the $t$th time slot by $S_{12}(t)$ and over $n$ time slots channel extension \textquotedblleft CSIT pattern\textquotedblright denoted by
$S_{12}^n=(S_{12}(1),...,S_{12}(n))$.

Furthermore, the fraction of time associated with the state of CSIT availability for the network denoted by $\lambda_{S}, S \in \{ P, D, N\}$ is given by 
\begin{equation}
\lambda_{S}=\frac{\sum_t\sum_i \mathbb{I}_S(S_i(t))}{k*n}
\end{equation}
where $\mathbb{I}$ denote the Indicator function and $k$ is the number of users, so that
\begin{equation}
\sum_S\lambda_{S}=1.
\end{equation}
and $\Lambda(\lambda_P, \lambda_D, \lambda_N )$ denotes the distribution of the fraction of time for the different states $\{P, D, N\}$ of the CSIT availability.
Let $r_{ij}(P)$ denote
the rate of $W_{ij}$ for a given transmission power $P$ where $
r_{ij}(P)= \frac{\log_2(|W_{ij}|)}{n}$ and $n$ is the number of
channel uses. The rate $r_{ij}(P)$ is achievable if there exists a
sequence of coding schemes such that the probability of error in
decoding $W_{ij}$ goes to zero as $n$ goes to infinity for all
$(i,j)$. The degrees of freedom region
$\mathcal{D}(\Lambda)$ is defined as the set of all achievable tuples
$(d_{11}, d_{12}, d_{21},d_{22})$, where $d_{ij}= \lim_{P
\rightarrow \infty} \frac{R_{ij}(P)}{\log_2{(P)}}$ is the degrees
of freedom for message $W_{ij}$. The sum degrees of freedom of
the network defined as:
\begin{equation}
\text{DoF}(\Lambda)= \max_{(d_{11},d_{12},d_{21},d_{22}) \in  \mathcal{D}(\Lambda)}
d_{11}+d_{12}+d_{21}+d_{22}
\end{equation}
Noteworthy, we could consider this system model as a general model for the availability of CSIT 
where the fixed perfect, delayed and no-CSIT are special cases included in this model for $\Lambda(1,0,0),\Lambda(0,1,0)\text{and} \Lambda(0,0,1)$ respectively. 
Finally, we could state the problem as to characterize the sum degrees of freedom $\text{DoF}(\Lambda)$ as a function of the distribution of the time fractions associated with the three states of CSIT availability for the X-Channel.

\section{ The achievability Scheme} \label{Scheme}

In this section, we present three illustrative cases for the
proposed achievable scheme in three different cases of CSIT
availability. In all these cases, we show that $5/4$ DoF is
achievable by sending $5$ different data symbols; $3$ for the first 
receiver and $2$ for the second one over four time slots.

Inspired by idea of Maddah-Ali and Tse in
\cite{maddah2012completely} which is to exploit the past 
received signals to create common signals to different receivers, 
hence improving DoF by broadcasting them to the receivers, we constructed our approach.
In the context
of distributed transmitters networks like X-channel, it may not be possible for transmitters 
to create common signals by reconstructing previously received signals with only delayed CSIT, 
since the common signal at one transmitter contains the signals of the other transmitters. 

Here, we developed a new transmission strategy based on creating signals of common interest for 
different receivers utilizing the synergy of alternating CSIT. 
 Specifically, exploiting the cooperation between delayed CSIT and perfect CSIT to perfectly reconstruct the \textquotedblleft interference\textquotedblright, signal of common interest, formerly received. 
Hence, transmitters could reconstruct the interference formerly received and defeat
the distributed nature of the X-channel.        

The proposed achievability scheme of $5/4$ for the SISO X-channel is 
performed in two phases over four time slots. 
The first phase is associated with the
delayed CSIT and divided into two sub-phases where each
sub-phase might consume one or two time slots. This phase is called 
\textquotedblleft interference
creation\textquotedblright, where
the transmitters greedily transmit their messages. As a result, the
receivers get linear combinations of their desired messages in
addition to interference. 
Totally, 
this phase needs three different delayed CSIT 
either distributed over three time slots, i.e., $S_{12}^3 = (ND,ND,DN)$
or combined over two time slots, i.e.,$S_{12}^2 = (DD,ND)$. 
On the other hand, the second phase is
associated with the perfect CSIT state and is called the
\textquotedblleft interference resurrection\textquotedblright phase. 
In this phase, transmitters reconstruct the old interference by exploiting the
delayed CSIT received in phase one and the perfect CSIT in the
second phase. Hence, after four time slots, one receiver has three
different linear combinations of its desired messages and only one
interference term received twice and the other receiver has two different 
linear combinations of its desired messages and three aligned interference terms. 
Noteworthy, in some cases the two phases can overlap over the four time slots.

Let $u^1_1$, $u^2_1$ and $u_2$ be three independent data symbol intended to
$R_1$ where $u^1_1$, $u^2_1$ transmitted from $T_1$ and $u_2$ transmitted from $T_2$.
Also, let
$v_1$ and $v_2$ be two independent data symbol intended to $R_2$
transmitted from $T_1$ and $T_2$, respectively. In the next
subsections, we show that we can reliably transmit the three symbols
$(u^1_1 ,u^1_2, u_2)$ to receiver 1 and the two symbols $(v_1 ,v_2)$ to
receiver 2 in four time slots in three different cases of alternating
CSIT.

\subsection{Combined creation and distributed resurrection}
Let us consider 
the alternating CSIT  pattern given by $S_{12}^4 = (DD,ND,PN,NN)$. 
Here, We have combined delayed $S_{12}(1) = (DD)$. Consequently, the interference
creation phase consumes only two time slots while the interference
resurrection phase can extend over two time slots. 
The proposed scheme is performed in two
separate phases as follows.

\textit{Phase one:} 
In the first time slot, each transmitter greedily
transmit two data symbols to the two receivers, i.e., $X_1(1)= u_1^1 + v_1$ and $X_2(1)=
u_2 + v_2$. As a result, the received signals
\begin{eqnarray}
Y_1(1)&\!\!\!=\!\!\!& h_{11}(1)u_1^1 + h_{12}(1)u_2+ h_{11}(1)v_1+h_{12}(1)v_2 \nonumber \\
&\!\!\!\equiv \!\!\!& L_1^1(u_1^1,u_2)+ I_1(v_1,v_2)\\
Y_2(1)&\!\!\!=\!\!\!&h_{21}(1)u_1^1 + h_{22}(1)u_2+ h_{21}(1)v_1+h_{22}(1)v_2 \nonumber \\
&\!\!\!\equiv\!\!\!& I_2^1(u_1^1,u_2)+ L_2^1(v_1,v_2)
\end{eqnarray}
where $L_i^j(x_1,x_2)$ denotes the $j$th linear combination of the
two messages $x_1$ and $x_2$ that are intended for receiver $R_i$
and $I_i^j(z_1,z_2)$ denotes the $j$th interference term for
receiver $R_i$ which is a function of the messages $z_1$ and $z_2$
that are not intended for this receiver.

Similarly, in the next time slot, $T_1$ transmits $u_1^2$ and $T_2$ transmits
$u_2$. The received signals at $R_1$ and $R_2$ are:
\begin{eqnarray}
Y_1(2)&\!\!\!=\!\!\!&h_{11}(2)u_1^2 + h_{12}(2)u_2 \equiv
L_1^2(u_1^2,u_2)\\
Y_2(2)&\!\!\!=\!\!\!&h_{21}(2)u_1^2 + h_{22}(2)u_2 \equiv
I_2^2(u_1^2,u_2)
\end{eqnarray}
Therefore, $R_1$ receives the second linear combination $L_1^2(u_1^2,u_2)$ of
its desired signals, while $R_2$ receives only interference
$I_2^2(u_1^2,u_2)$. 
 
\textit{Phase two:} This phase consists of two time slots
where in each time slot the transmitted signals are designed such
that the interference is resurrected at one receiver while the
second receiver receives a new linear combination of its desired
messages. Note that now the transmitters are aware of the CSIT of
the previous time slot, i.e., $T_1$ knows $h_{11}(1)$, $h_{21}(1)$ and
$h_{21}(2)$ while $T_2$ knows $h_{12}(1)$, $h_{22}(1)$ and $h_{22}(2)$. 
Also, at $t=3$, the channels to the first receiver are known perfectly
and instantaneously at the two transmitters, i.e., $T_1$ knows
$h_{11}(3)$ and $T_2$ knows $h_{12}(3)$. As a result, the first
time slot in this phase is dedicated to resurrecting the
interference $I_1(v_1,v_2)$ received by $R_1$ in the first time
slot. The transmitted signals of $T_1$ and $T_2$ are: 
\begin{eqnarray}
X_1(3)&\!\!\!=\!\!\!&h_{11}^{-1}(3)h_{11}(1)v_1\\
X_2(3)&\!\!\!=\!\!\!&h_{12}^{-1}(3)h_{12}(1)v_2
\end{eqnarray}
and the received signals at $R_1$ and $R_2$ are:
\begin{eqnarray}
Y_1(3)&\!\!\!=\!\!\!& h_{11}(1)v_1+h_{12}(1)v_2 \equiv
I_1(v_1,v_2)\\
Y_2(3)&\!\!\!=\!\!\!&h_{21}(3)h_{11}^{-1}(3)h_{11}(1)v_1+h_{22}(3)h_{12}^{-1}(3)h_{12}(1)v_2 \nonumber \\
&\!\!\!\equiv \!\!\!&L_2^2(v_1,v_2)
\end{eqnarray}
Hence, at the end of sub-phase one, $R_1$ has received
$I_1(v_1,v_2)$, the interference received in the first time
slot, and $R_2$ has received a new linear combination
$L_2^2(v_1,v_2)$.

In the second sub-phase; the fourth time slot, the transmitted signals are designed to
resurrect the interference received by $R_2$ in the first and second time
slots and provide a new linear combination of the desired messages
to $R_1$. The transmitted signals of $T_1$ and $T_2$ are given by:
\small
\begin{eqnarray}
X_1(4)&\!\!\!=\!\!\!&h_{22}(2)h_{21}(1)u_1^1-h_{22}(1)h_{21}(2)u_1^2\\
X_2(4)&\!\!\!=\!\!\!&0
\end{eqnarray}
\normalsize
The received signals at $R_1$ and $R_2$ are given by:
\small
\begin{eqnarray}
Y_1(4)&\!\!\!=\!\!\!&h_{11}(4)[(h_{22}(2)h_{21}(1)u_1^1
- h_{22}(1)h_{21}(2)u_1^2]\nonumber\\&\!\!\!\equiv
\!\!\!&
L_1^3(u_1^1,u_1^2) \\
Y_2(4)&\!\!\!=\!\!\!&h_{21}(4)[(h_{22}(2)h_{21}(1)u_1^1
- h_{22}(1)h_{21}(2)u_1^2]\nonumber\\&\!\!\!\equiv
\!\!\!&h_{21}(4)[h_{22}(2)I_2^1(u_1^1,u_2)- h_{22}(1)I_2^2(u_1^2,u_2)]
\end{eqnarray}
\normalsize

After the fourth time slot, the two receivers $R_1$ and $R_2$ have
enough information to decode their intended messages. In
particular, $R_1$ have access to three different equations in $u_1^1$
$u_1^2$ and $u_2$. The first one is obtained by subtracting $Y_1(3)$,
from $Y_1(1)$ to cancel out the interference, the second
and the third equations are $Y_1(2)$ and $Y_1(4)$ by themselves as they are received without
interference. Similarly, $R_2$ forms its first equation by
subtracting $h_{21}^{-1}(4)Y_2(4)$ from $(h_{22}(2)Y_2(1)- h_{22}(1)Y_2(2))$ to cancel out the interference while the second equation is $Y_2(3)$.

\subsection{Distributed creation and combined resurrection}
Let us consider the 2-user SISO X-channel with alternating CSIT
given by $S_{12}^4 = (ND,ND,DN,PN)$. Here, we have the delayed CSIT
are distributed over three time slots. Consequently, the interference
creation phase consumes three time slots while the interference
resurrection phase could be executed over only one time slots. 
The transmission strategy of 
this case are similar to case 1 but with minor modification 
concisely clarified below and the details of
the transmission strategy are omitted for brevity.
  
\textit{Phase one:} the first and second time slots of this phase are dedicated to
the first receiver where the transmitters transmit two different 
linear combinations form the desired messages.
The third time slot is dedicated to
the second receiver where $T_1$ transmits $v_1$ and $T_2$ transmits
$v_2$.  

\textit{Phase two}: This phase includes only the fourth time slot where the
transmitters resurrect the formerly received interference terms
$I_1(v_1,v_2)$, $I_2^1(u_1^1,u_2)$ and $I_2^2(u_1^2,u_2)$, while providing new linear
combinations of the desired messages to the two receivers. Therefore, the transmitted signals from $T_1$ and $T_2$ is given by
\begin{eqnarray}
X_1(4)&\!\!\!=\!\!\!&h_{22}(2)h_{21}(1)u_1^1-h_{22}(1)h_{21}(2)u_2^2\nonumber \\
&\!\!\!+ \!\!\!&h_{11}^{-1}(4)h_{11}(3)v_1\\
X_2(4)&\!\!\!=\!\!\!&h_{12}^{-1}(4)h_{12}(3)v_2
\end{eqnarray}
by the end of the fourth time slot, the two receivers $R_1$ and $R_2$ have
enough information to decode their intended messages.
For example, $R_1$
subtracts $Y_1(3)$ from $Y_1(4)$ to obtain the third equation in $u_1^1$ and $u_2^2$ while the first and second equations are $Y_1(1)$ and $Y_1(2)$ by themselves.

\subsection{Distributed creation and distributed resurrection}
Let us consider a 2-user SISO X-channel with CSIT pattern given by $S_{12}^4 = (ND,DN,PD,NN)$.
Unlike the aforementioned cases, we have an overlap between the two phases in the third time slot where $S_{12}(3) = (PD)$ occurs. Consequently, the proposed scheme is
performed in two overlapping phases as follows.

\textit{Time slot 1:} The first sub-phase of phase one begins at
$t=1$, and is dedicated to transmitting the desired messages of
$R_1$, i.e., $T_1$ transmits $u_1^1$ while $T_2$ transmits $u_2$.
Therefore, $R_1$ receives the first linear combination $L_1^1(u_1^1,u_2)$ of its desired signals, while $R_2$ receives only interference $I_2^1(u_1^1,u_2)$.

\textit{Time slot 2}: The second sub-phase of phase one occurs at
$t=2$, and is dedicated to transmitting the desired messages of
$R_2$, i.e., $T_1$ transmits $v_1$ while $T_2$ transmits $v_2$.
Therefore, $R_2$ receives linear combination $L_2^1(v_1,v_2)$ of its desired signals, while $R_1$ receives only interference $I_1(v_1,v_2)$.

\textit{Time slot 3}: At $t=3$ the overlap occurs between the two
phases. In particular, the second time slot of sub-phase one of phase one
and sub-phase one of phase two begin simultaneously. In this time slot,
sub-phase one of phase one creates interference at $R_2$ with
while sub-phase one of phase two is designed to resurrect the
interference term $I_1(v_1,v_2)$. The transmitted signals are
given by:
\begin{eqnarray}
X_1(3)&\!\!\!=\!\!\!&u_1^2+h_{11}^{-1}(3)h_{11}(2)v_1 \\
X_2(3)&\!\!\!=\!\!\!&u_2+h_{12}^{-1}(3)h_{12}(2)v_2
\end{eqnarray}
and the corresponding received signals are given by:
\begin{eqnarray}
Y_1(3)&\!\!\!=\!\!\!&h_{11}(3)u_1^2 + h_{12}(3)u_2 +h_{11}(2)v_1 + h_{12}(2)v_2 \nonumber \\
&\!\!\!\equiv\!\!\!&  L_1^2(u_1^2,u_2)+I_1(v_1,v_2)\\
Y_2(3)&\!\!\!=\!\!\!& h_{21}(3)h_{11}^{-1}(3)h_{11}(2)v_1+h_{22}(3)h_{12}^{-1}(3)h_{12}(2)v_2 \nonumber \\
&\!\!\!+\!\!\!&h_{21}(3)u_1^2 + h_{22}(3)u_2\nonumber \\
&\!\!\!\equiv\!\!\!& L_2^2(v_1,v_2)+I_2^2(u_1^2,u_2)
\end{eqnarray}
Therefore, $R_2$ receives a new linear combination
$L_2^2(v_1,v_2)$ of its desired signals and an interference term
$I_2(u_1^2,u_2)$ as a by-product of the overlap, while $R_1$
receives the old interference $I_1(v_1,v_2)$ and the second linear
combination $L_1^2(u_1^2,u_2)$ of its desired signals.

\textit{Time slot 4}: The details of
the transmission strategy are omitted for brevity 
as it is almost similar to the fourth time slot of case 1.
Finally, the two receivers $R_1$ and $R_2$ have enough information
to decode their intended messages.
 Similarly, $R_2$
its first equation is $Y_2(2)$ while forming its second equation
by subtracting $h_{21}^{-1}(4)Y_2(4)$ from $(h_{22}(1)Y_2(3)- h_{22}(3)Y_2(1))$ to cancel out the
interference.

Noteworthy, this scheme could be used not only with the aforementioned cases but also with all the synergistic alternating CSIT patterns listed in Table 1. Also, It could be used with all the mirrored copy  of the synergistic alternating CSIT patterns listed in Table 1 but with minor modification in the two phases where the two sub-phases in each phase swap their dedications from
$R_1$ to $R_2$ and vise versa.

\section{Main Results and Discussion}
To the best of our knowledge, there are only three achievability results 
and fundamental bounds on the DoF for the 2-user SISO X-channel. 
In particular, $4/3$ DoF for 2-user SISO X-channel with perfect \cite{5208535}, $6/5$ DoF \cite {ghasemi2011degrees} and
unity DoF \cite{6205390} for the same channel with delayed CSIT and no-CSIT, respectively. The poverty 
in the achievablility of DoF results in the context of X-channel
compared with the other wireless channels 
encourage us to introduce new 
achievable results and hence $5/4$ DoF hits. 

In this section, we discuss CSIT alternation patterns that can
provide synergistic gain for the DoF of the two-user SISO X-channel.
We note that the aforementioned examples in Section
\ref{Scheme} present the synergistic alternating CSIT patterns among
four-symbol channel extension CSIT patterns of $\Lambda(1/8,3/8,1/2)$
sufficient to achieve $5/4$ DoF. 

\textit{Theorem 1:} The two-user SISO X-channel with synergistic alternating CSIT
 and the associated distribution of the fraction of time for the different
 CSIT availability states $\{P, D, N\}$ is $\Lambda(1/8,3/8,1/2)$, 
 can achieve $5/4$ degrees of freedom almost surely.  
\begin{proof}[\textit{Proof:}\\]

 \textit{A.  }  \textit{Achievability:}\\
The achievability is provided in section \ref{Scheme}, and utilizes a simple linear coding scheme to
achieve $5/4$. The coding scheme is based on the idea of reconstructing the interference formerly received exploiting the synergy of delayed CSIT followed by perfect CSIT to realize interference alignment at receivers. 

\textit{B.  }  \textit{Distribution:}\\
Here we show that the first requirement for Theorem 1; the associated distribution of the fraction of time for the CSIT availability states $\{P, D, N\}$ is $\Lambda(1/8,3/8,1/2)$, contains sufficient channel knowledge that allows the transmitters to steer their signals to perfectly align interference at the receivers. As mentioned before to achieve $5/4$ DoF we send $5$ different symbols $3$ for one receiver and $2$ for the other one over $4$-symbol channel extension. 
$\Lambda(1/8,3/8,1/2)$ contains only $1$ perfect CSIT, $3$ delayed CSIT and $4$ no-CSIT over $4$ time slots. 
Three delayed CSIT are sufficient in the interference creation phase to create $3$ different constructable interference terms and as a by-product $2$ different linear combination of desired symbols to one receiver and only one to the other one.
Noteworthy, each delayed CSIT could be associated with no-CSIT in the interference creation phase as receiving interference at one receiver imply receiving desired signal at the other receiver. Finally, only  one perfect CSIT is necessary to broadcast one old interference term reconstructed from distributed transmitters, i.e. $I_1(v_1,v_2)$, and only one no-CSIT to broadcast one old interference term reconstructed from one transmitter, i.e. $I_2(u_1^1,u_1^2)$.

\textit{C.  }  \textit{Synergistic alternation:} \\
Although $\Lambda(1/8,3/8,1/2)$ contains sufficient channel knowledge to enable interference alignment, it is not sufficient by itself to do that. Hence, the synergistic alternation in CSIT arises to be a complementary condition to $\Lambda(1/8,3/8,1/2)$ to be a sufficient condition to realize the interference alignment and achieve $5/4$ DoF. There are three conditions for the synergistic alternation of CSIT for $\Lambda(1/8,3/8,1/2)$. The first condition is no delayed CSIT at the last time slot as it will be  degraded to no-CSIT. The second condition is the delayed CSIT should followed by perfect CSIT 
for the same receiver. The third condition is the $S_{12} \in \{NN\}$ is prohibited in the interference creation phase. The first and second conditions yield six possible minimum states for the CSIT of the channel to the receiver interested in receiving $3$ symbols over four time slots; \small $S_i^4 \in \{\text{(N,N,D,P),(N,D,N,P),(D,N,N,P),(N,D,P,N),(D,N,P,N),(D,P,N,N)}\}$\normalsize, and three possible minimum states for the CSIT of the channel to the other receiver;  
$S_i^4 \in \{\text{(D,D,N,N), (D,N,D,N), (N,D,D,N)}\}$. As a result we have 18 possible combinations for the CSIT of the two-user channel. Eleven of these 18 combinations, satisfy the third condition and are listed as the first $11$ entries in Table 1 and the other excluded by the third condition are the dissociative ones .\\
\renewcommand{\qedsymbol}{$\blacksquare$}
\end{proof}                                                                            
\begin{table}[!ht]
\begin{tabular}{|l|p{.85 cm}||l|p{.85 cm}|}
  \hline
   CSIT pattern & Case & CSIT pattern &  Case  \\
  \hline
 $(DD,ND,PN,NN)$ &  1& $(ND,DN,ND,PN)$  &  2\\
 $(ND,DD,PN,NN)$ &  1& $(DN,ND,ND,PN)$  &  2\\
 $(ND,DD,NN,PN)$ &  1& $(ND,DN,PD,NN)$  &  3\\
 $(DD,ND,NN,PN)$ &  1& $(DN,ND,PD,NN)$  &  3\\
 $(DD,PN,ND,NN)$ &  1& $(DN,PD,ND,NN)$  &  3\\
 $(ND,ND,DN,PN)$ &  2&                  &   \\
  \hline
\end{tabular}\\
\caption {the synergistic alternating CSIT patterns for $\Lambda(1/8,3/8,1/2)$ }
\end{table}


\textit{Remark 1:} \textbf{[Redundant Knowledge]}
The first condition for achievability of $5/4$ DoF in Theorem 1 clarify that it requires
alternating CSIT with $\Lambda(1/8,3/8,1/2)$. Definitely, any synergistic alternating CSIT pattern with higher distribution $\Lambda(\lambda_P \geq 1/8, \lambda_D \geq 3/8, \lambda_N \leq 1/2)$ can achieve $5/4$ DoF for SISO X-channel but with much redundant channel knowledge. Although we showed that $\Lambda(1/8,3/8,1/2)$ provide zero redundant channel knowledge for the achievable scheme in section \ref{Scheme}, it still unproven that $\Lambda(1/8,3/8,1/2)$ is the minimum distribution sufficient for synergistic alternating pattern to achieve $5/4$ or not.   

\textit{Remark 2:} \textbf{[Synergistic alternation pattern ]} The CSIT alternation pattern listed in Table I are called synergistic patterns as they could interact together in a cooperative way to provide a DoF gain strictly greater that their individual DoF for the same network. For alternating CSIT with $\Lambda(1/8,3/8,1/2)$, if there is no interaction between the different CSIT states over four time slots,    
the DoF that could be achieved are $\frac{1}{8}
\frac{4}{3} + \frac{3}{8} \frac{6}{5} + \frac{1}{2}= \frac{67}{60}$ which is
 strictly lower than $5/4$. \\

\textit{Remark 3:} \textbf{[Blind creation]}
The third condition for the synergistic CSIT alternation patterns of $\Lambda(1/8,3/8,1/2)$ excludes seven 
CSIT alternation patterns where the synergy in alternation could not achieve $5/4$. The main common factor of the seven dissociative patterns is the existence of combined no-CSIT, i.e. $S_{12}= NN$, at the interference creation phase, i.e. $S_i^4 =(NN, ND, DD, PN)$. As mentioned before, the strategy of interference creation phase is to create interference for the receiver who can provide the transmitters with CSIT either perfect or delayed to enable the transmitters to reconstruct the interference in the interference resurrection phase. Then, combined no-CSIT is useless at interference creation phase as it provides nothing to transmitters and hence the transmitters blindly create interference.

\section{Conclusion}
We obtained new achievable results on the degrees of freedom of
 the X-channel with alternating CSIT under certain distribution of CSIT availability states. 
The achieved DoF under alternating CSIT assumption are strictly greater
than the DoF of the same channel under delayed and no-CSIT assumptions.
By developing new transmission strategy, the usefulness of perfect CSIT
when available following delayed CSI has been illustrated.

%

\bibliographystyle
{IEEEtran}
\bibliography{IEEEabrv,Nulls}

\begin{thebibliography}{10}
\providecommand{\url}[1]{#1}
\csname url@samestyle\endcsname
\providecommand{\newblock}{\relax}
\providecommand{\bibinfo}[2]{#2}
\providecommand{\BIBentrySTDinterwordspacing}{\spaceskip=0pt\relax}
\providecommand{\BIBentryALTinterwordstretchfactor}{4}
\providecommand{\BIBentryALTinterwordspacing}{\spaceskip=\fontdimen2\font plus
\BIBentryALTinterwordstretchfactor\fontdimen3\font minus
  \fontdimen4\font\relax}
\providecommand{\BIBforeignlanguage}[2]{{%
\expandafter\ifx\csname l@#1\endcsname\relax
\typeout{** WARNING: IEEEtran.bst: No hyphenation pattern has been}%
\typeout{** loaded for the language `#1'. Using the pattern for}%
\typeout{** the default language instead.}%
\else
\language=\csname l@#1\endcsname
\fi
#2}}
\providecommand{\BIBdecl}{\relax}
\BIBdecl

\bibitem{maddah2008communication}
M.~A. Maddah-Ali, A.~S. Motahari, and A.~K. Khandani, ``Communication over
  {MIMO} {X} channels: Interference alignment, decomposition, and performance
  analysis,'' \emph{IEEE Transactions on Information Theory}, vol.~54, no.~8,
  pp. 3457--3470, 2008.

\bibitem{4567443}
V.~Cadambe and S.~Jafar, ``Interference alignment and degrees of freedom of the
  {K}-user interference channel,'' \emph{IEEE Transactions on Information
  Theory}, vol.~54, no.~8, pp. 3425--3441, 2008.

\bibitem{5208535}
------, ``Interference alignment and the degrees of freedom of wireless {X}
  networks,'' \emph{IEEE Transactions on Information Theory}, vol.~55, no.~9,
  pp. 3893--3908, 2009.

\bibitem{4418479}
S.~Jafar and S.~Shamai, ``Degrees of freedom region of the {MIMO} {X}
  channel,'' \emph{IEEE Transactions on Information Theory}, vol.~54, no.~1,
  pp. 151--170, 2008.

\bibitem{6205390}
C.~Vaze and M.~Varanasi, ``The degree-of-freedom regions of {MIMO} broadcast,
  interference, and cognitive radio channels with no {CSIT},'' \emph{IEEE
  Transactions on Information Theory}, vol.~58, no.~8, pp. 5354--5374, 2012.

\bibitem{maddah2012completely}
M.~A. Maddah-Ali and D.~Tse, ``Completely stale transmitter channel state
  information is still very useful,'' \emph{IEEE Transactions on Information
  Theory}, vol.~58, no.~7, pp. 4418--4431, 2012.

\bibitem{6471826}
R.~Tandon, S.~Jafar, S.~Shamai~Shitz, and H.~Poor, ``On the synergistic
  benefits of alternating csit for the miso broadcast channel,'' \emph{IEEE
  Transactions on Information Theory}, vol.~59, no.~7, pp. 4106--4128, 2013.

\bibitem{ghasemi2011degrees}
A.~Ghasemi, A.~S. Motahari, and A.~K. Khandani, ``On the degrees of freedom of
  {X} channel with delayed {CSIT},'' in \emph{proceedings IEEE International
  Symposium on Information Theory (ISIT) 2011}.

\bibitem{Delayed_CSIT}
S.~Lashgari, A.~S. Avestimehr, and C.~Suh, ``Linear degrees of freedom of the
  x-channel with delayed csit,'' \emph{CoRR}, vol. abs/1309.0799, 2013.

\bibitem{Wagdy}
A.~Wagdy, A.~El-Keyi, T.~Khattab, and M.~Nafie, ``A dof-optimal scheme for the
  two-user x-channel with synergistic alternating csit,'' \emph{CoRR}, vol.
  abs/1404.6348, 2014.

\end{thebibliography}

\end{document}